\documentclass[aps,prl,preprint,altaffilletter,12pt,showpacs]{revtex4}
\usepackage{graphicx}
\usepackage{amssymb}
\usepackage{hyperref}

\begin{document}
\title{Viscoelastic Suppression of Gravity-Driven Counterflow Instability}
\author{P.~Beiersdorfer}
\author{D.~Layne}
\author{E.~W.~Magee}
\affiliation{Lawrence Livermore National Laboratory, Livermore, CA 94550}
\author{J.~I. Katz}
\affiliation{Department of Physics and McDonnell Center for the Space
Sciences, Washington University, St.~Louis, Mo. 63130}
\altaffiliation{Lawrence Livermore National Laboratory, Livermore, CA 94550}
\email{katz@wuphys.wustl.edu}
\date{\today}

\begin{abstract}
Attempts to achieve ``top-kill'' of flowing oil wells by pumping dense
drilling ``muds'', {\it i.e.\/}, slurries of dense minerals, from above will
fail if the Kelvin-Helmholtz instability in the gravity driven counterflow
produces turbulence that breaks up the denser fluid into small droplets. 
Here we estimate the droplet size to be sub-mm for fast flows and suggest
the addition of a shear-thickening or viscoelastic polymer to suppress
turbulence.  We find in laboratory experiments a variety of new physical
effects for a viscoelastic shear-thickening liquid in a gravity driven
counterstreaming flow.  There is a progression from droplet formation to
complete turbulence suppression at the relevant high velocities.  Thick
descending columns show a viscoelastic analogue of the viscous buckling
instability.  Thinner streams form structures resembling globules on a
looping filament.
\end{abstract}


\maketitle

Dense fluids, {\it i.e.\/}, mineral suspensions called ``mud''
\cite{DG88,VD00}, are introduced into oil wells to provide hydrostatic
pressure to offset hydrocarbon (oil and gas) fluid pressure in deep
formations, stopping upward flow and reducing the fluid pressure at the
surface to near ambient.  In a procedure known as ``top kill'' these muds
are introduced at the top of the well.  After pressure-driven injection they
descend, filling it, in a gravity driven flow.  

If hydrocarbon is flowing upward in the well there is a counterflow between
the upwelling hydrocarbon and the descending mud.  Successful top-kill
requires that the mud descend despite this counterflow.  However, upwelling
at speeds $\gtrsim 1$ m/s, as in the uncontrolled Macondo well ``blow-out''
in the Gulf of Mexico in 2010, may lead to a Kelvin-Helmholtz instability
\cite{B67}.
%
In an attempt at top-kill this may disrupt the
coherent downward stream of the mud by breaking it into small packets or
droplets, and may explain the failure of the attempted top-kill of Macondo.

In order for a denser fluid to descend through an upwelling vertical column
of lighter fluid it must not break up into packets or droplets whose
settling velocity is less than the upwelling velocity.  In vertical
counterflows at high Reynolds number the Kelvin-Helmholtz instability grows
and progresses rapidly to turbulence. 
This is in contrast to the case of
horizontal flows in the stratified atmosphere or ocean, in which gravity and
the vertical density gradient act to stabilize the instability.

If instability is not suppressed, the dense fluid
introduced at the top of the column will be dispersed into small droplets
that are spat out with the upwelling flow.  In a ``blown-out'' oil well the
consequence would be the failure of the dense fluid to accumulate to a depth
sufficient to provide the hydrostatic pressure head necessary to ``kill''
(suppress the entry of hydrocarbon into) the well.

The purpose of this paper is to report experiments demonstrating that a
surrogate mud is not dispersed by oil-mud counterflow when the
mud consists of a shear-thickening viscoelastic suspension.  Two-phase flows
exhibit many complex phenomena \cite{L99,K03}.  There appear to be no
previous studies of two-phase flows in which one of the fluids is
viscoelastic.

Conventional drilling muds are typically Bingham plastics that flow as
shear-thinning (pseudo-plastic) fluids above a small elastic yield stress
\cite{DG88,VD00,S96,LP05}.  This permits the mud to remove cuttings and
debris and to keep them suspended during interruptions in drilling, while
minimizing the required pumping pressure and power.  In the following, we
show that such fluids are likely to be dispersed into small pockets or
droplets and carried out of the well bore by a rapid counterflow.

Because the theoretical estimates of this paper imply that instability and
turbulent mixing will be severe with conventional muds, we suggest that the
use of a shear-thickening mud that becomes elastic at high strain rates may
enable top-kill of a rapidly flowing blown-out well.  We demonstrate in
experiments with viscoelastic shear-thickening fluids complete suppression
of the Kelvin-Helmholtz instability at flow velocities and Reynolds numbers
approaching those of the blown-out Macondo well.  

At lower flow rates thin filaments of immiscible fluid break up into
droplets (the Plateau-Rayleigh instability \cite{dG02}) under the influence
of interfacial tension. 
We also demonstrate suppression of this instability in thin jets of
viscoelastic fluid as elasticity prevents the rupture of the jet.  In its
place, we find novel ``globule and filament'' phenomenology.

Nonlinear instabilities and turbulence present formidable problems.  We 
estimate the size of the droplets produced by turbulence, if it occurs, by
assuming a Kolmogorov turbulent cascade \cite{TL72} driven by the
Kelvin-Helmholtz instability at an outer scale $L$ and velocity $U$.  $L$ is
taken as the diameter of the descending mud column (a fraction, determined
by the mud flow rate, of the well bore diameter).  $U$ is the velocity
difference between oil and mud, and must exceed the upwelling velocity of
the oil for the mud to descend at all.

In an oil well $L$ and $U$ vary with depth.  Turbulence is most intense and
most effectively disperses the dense mud at the bottom of the well where $L$
is smallest and $U$ greatest, but these parameters do not vary by large
factors, in part because in the upper, larger diameter, portions of the
bore only an annulus may need to be filled \cite{Macondo}.

Adopting $U = 3.7$~m/s (corresponding to Macondo's estimated \cite{CT10}
uncontrolled flow of 50,000 barrels/day (92 l/s) through the $0.18$ m
diameter bore at the bottom of the well \cite{Macondo}) and $L = 0.09$ m
(considering a mud column half the bore diameter), and taking a
representative kinematic viscosity of crude oil of $10^{-5}$ m$^2$/s
\cite{VD97} leads to a Reynolds number of roughly 40,000.  Turbulence would
be expected, but a well developed inertial subrange \cite{TL72} would not
be.  The problem is complicated by the presence of two dynamically
interacting fluids with very different properties.  For lack of a better
theory, we use the Kolmogorov spectrum in order to estimate the size of the
resulting spatial structure.

If the fluids are miscible, the inner Kolmogorov scale (the smallest length
scale on which the fluid is turbulent) \cite{TL72} provides an estimate of
the scale of compositional heterogeneity:
\begin{equation}
r_{visc} \approx \nu^{3/4} \epsilon^{-1/4} \approx \nu^{3/4} L^{1/4}
U^{-3/4},
\label{rvisc}
\end{equation}
where $\epsilon \approx U^3/L$ is the specific energy dissipation rate and
$\nu$ is the kinematic viscosity.  The shear rate is high for $r >
r_{visc}$, but decays $\propto \exp{(- 1.5 (r/r_{visc})^{-4/3})}$ for $r <
r_{visc}$ \cite{TL72}.  Structure on the scale $r_{visc}$ develops in a time
$\ll L/U$, but no finer spatial structure is expected.  For miscible fluids
the spatial structure does not take the form of spherical droplets because
there is no interfacial energy acting to minimize the area of the boundary;
we refer instead to ``packets'' of denser fluid.

The formation of a packet or droplet of one fluid surrounded by the other
requires shear flow on the heterogeneity scale in both fluids.  We adopt
for the viscosity $\nu$ that of crude oil.  For a comparatively inviscid,
{\it i.e.\/}, not shear thickening, mud, the characteristic size of the 
heterogeneity is $r_{visc} \sim 0.05$ mm.  These denser mud volumes have a
Stokesian descent (settling) velocity (for a density difference of 1
gm/cm$^3$) $v_{descent} < 1$ cm/s, far less than the upward velocity of the
fluid in which they are immersed.   Even at the
inner Kolmogorov scale the Reynolds stress $\rho u_k^2 \sim \rho
(\epsilon/k)^{2/3} \sim \rho (\epsilon \nu)^{1/2}$ far exceeds the Bingham
yield stress $\lesssim 10$ Pa of typical drilling muds \cite{DG88} so that
these muds behave essentially as shear-thinning (pseudoplastic) fluids.

If the fluids are immiscible, as is the case with water-based muds, then
interfacial energy further limits the droplet size,
although it does not change the Kolmogorov inner scale of the turbulence
within the homogeneous fluid.  Equating the turbulent kinetic energy $\rho
u_k^2/2$ on a wave number scale $k$ in the volume of a droplet of radius
$r_{surf}$ to its interfacial energy yields
\begin{equation}
{4 \pi r_{surf}^3 \over 3} {\rho \epsilon^{2/3} \over 2 k^{2/3}} \approx 
4 \pi \sigma r_{surf}^2;
\end{equation}
this is equivalent to the condition that the Weber number We $\equiv \rho
u_k^2 r_{surf}/\sigma = {\cal O}(1)$, where $u_k \approx
(\epsilon/k)^{1/3}$.  Taking $k = 1/r_{surf}$,
\begin{equation}
r_{surf} \approx \left({6 \sigma \over \rho}\right)^{3/5} {L^{2/5} \over
U^{6/5}}.
\label{rsurf}
\end{equation}

The oil-water interfacial energy $\sigma \approx 0.025$ N/m.  The
characteristic size of the droplets produced $r_{surf} \sim 0.7$ mm.  The
actual droplet sizes will be the greater of $r_{visc}$ and $r_{surf}$; 
$r_{surf} > r_{visc}$ for the assumed parameters.  The turbulent Reynolds
stress at the smallest turbulent scale exceeds the typical Bingham yield
stress (by an even larger factor than for miscible fluids if evaluated
at the scale $k = 1/r_{surf}$).
The Stokesian descent velocity $v_{descent} \sim 10$ cm/s, again less than
the upward velocity of the fluid in which the droplets are immersed (finite
Reynolds number effects reduce the settling speed by a factor $\gtrsim 1$).
The Weber number obtained from the settling speed We$_{descent} \equiv
\rho v_{descent}^2 r_{surf}/\sigma \sim 0.3$, justifying the assumption that
the droplets remain nearly spherical and intact.

For parameters over a broad range approximating the best estimates of those
of the Macondo blowout, dense fluid introduced at the top of the
well would not have sunk to the bottom, but would have been swept out with
the escaping crude oil.  This explains the failure of top kill in that well.
Similarly, had bottom-kill, {\it i.e.\/}, the introduction of dense fluid at
the bottom of the well, been attempted while the well was flowing at 
$\approx 3$ m/s, it would also have failed.


It is possible to suppress instability and thereby to avoid dispersion of
the mud into small droplets and its sweep-up by the counterflowing oil by
adding to it a dilatant polymer with shear-thickening and viscoelastic
properties.  Before instability could grow to an amplitude sufficient to
disperse the mud, the mud would have become very viscous or even elastic in
the counterflow shear layer where dispersion would otherwise occur.
Kelvin-Helmholtz instability is slowed by viscosity \cite{B67}, and can be
stabilized by elasticity.  If instability occurred, once its growth produced
a sufficient shear rate the mud would have become viscoelastic, suppressing
further growth and preventing its dispersal.  The resulting two-phase flow
is difficult to predict and likely to be complex \cite{L99,K03}.  One of the
purposes of our experiments was to study these phenomena. 
 
Corn starch-water emulsions are the classic shear-thickening viscoelastic
fluid \cite{M04,F08}.  Viscoelastic behavior is observed for mass ratios
$\rho^*$ of corn starch to water in the range 1.1--1.7 (within this range
there is little dependence on $\rho^*$).  At low shear rates they shear-thin
from a viscosity of about 100 Pa-s at 0.02/s to about 10 Pa-s at 2--5/s.
Above this shear rate they abruptly shear-thicken in a jamming transition to
a dynamic viscosity $\eta_{jam} \gtrsim 1000$ Pa-s.  At these higher shear
rates the emulsion also has an elastic response that is difficult to
quantify, but that is manifested in the trampoline-like behavior of pools of
corn starch-water emulsions.

The Kolmogorov inner scale turbulent shear rate $\sqrt{\epsilon/\nu} \approx
U^{3/2} / \sqrt{L \nu} \sim 8000$/s would far exceed that required for shear
thickening and elasticity, so that the addition of corn starch to the mud
would be expected to prevent its dispersal.  Even the outer scale shear rate
$U/L \sim 40$/s at the mud-crude oil boundary is sufficient to put the mud
in the shear-thickened viscoelastic regime and to suppress the
Kelvin-Helmholtz instability.

A column of dense viscoelastic fluid is thus predicted to remain coherent,
with its flow described by a (shear-thickened) Reynolds number $\rho U L
/\eta_{jam} = {\cal O} (1)$, sufficient to slow the Kelvin-Helmholtz
instability and to preclude turbulence.  Its descent would be retarded only
by the viscous drag of the surrounding (Newtonian) lighter oil.  As the
column accelerates downward under the influence of its negative buoyancy it
would stretch into a thin filament, limited by its viscoelasticity.

In order to test these hypotheses we filled a transparent column 1.6 m tall
and 63 mm in internal diameter with a transparent light mineral oil
\cite{Multitherm} of density 790 kg/m$^3$ and viscosity 6.4 mPa-s.  
Although the aspect ratio (depth to diameter) of the column was much less
than that of a real oil well (several km deep), the instability that would
disperse the mud would be expected to occur in a comparatively short (${\cal
O}$(10) diameters) length of the column, were it to occur at all.  The
subsequent flow would be expected to be gravity-driven settling, slow if
the mud were dispersed but rapid if a coherent slug.  Hence these
experiments are applicable to any column or well with aspect ratio $\gg 1$.  

We released 0.15 l of water (with a dye added for visibility) from a funnel
at a mean flow rate of 0.11 l/s, and observed vigorous turbulence and
dispersal of the stream into small (radius $\sim 1$ mm) droplets.  The
corresponding Reynolds number Re, based on the diameter of the water stream
and the viscosity of the oil, was about 2000 at the speed (1.15 m/s) of
entry of the water into the oil, so turbulence was expected \cite{BS63}  (Re
based on the viscosity of water was $\sim 10^4$).  Aside from the volumetric
flow rate and column diameter, the parameters, including the Atwood number
At $\equiv (\rho_{water} - \rho_{oil})/ (\rho_{water} + \rho_{oil})$, were
comparable to those of drilling mud in a light crude oil; the predicted
droplet radius $r_{surf} \approx 0.7$ mm.  Unlike drilling muds, our fluids
contained no surfactant, so most of these small globules rapidly coalesced
once the turbulence decayed, and descended in our static oil column
(Fig.~1(a)).
\begin{figure}
\begin{center}
\includegraphics{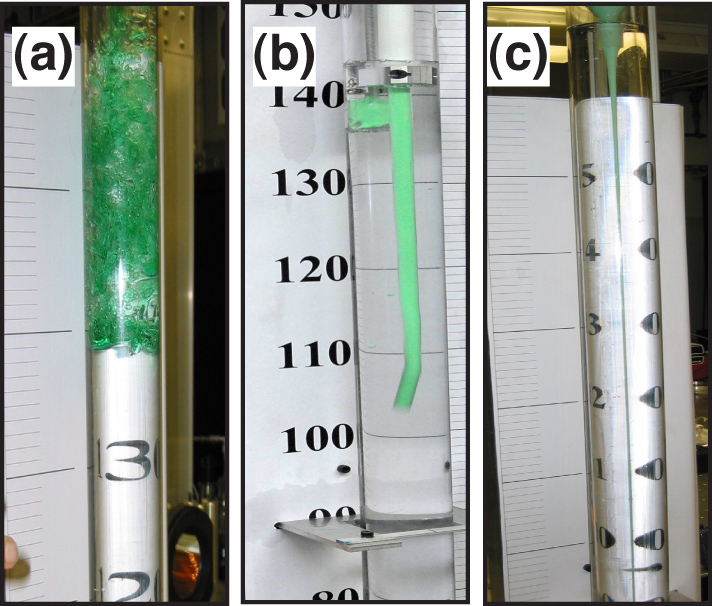}
\caption{Left shows turbulent breakup following Kelvin-Helmholtz instability
of descending water in oil column at 1.15 m/s and 0.11 l/s; center shows
descent of a coherent slug of viscoelastic suspension of corn starch in
water (mass ratio $\rho^*$ = 1.3:1) driven by a plunger at 0.56 m/s and 0.11
l/s with stabilization of the instability; right shows stable descent of the
same suspension at 1--2 ml/s following passage of leading edge.  Vertical
divisions are cm and tens of cm.}
\end{center}
\end{figure}

We then repeated the experiment using a strongly viscoelastic aqueous
suspension ($\rho_{susp} = 1.30$ gm/cm$^3$ and At = 0.24; these values are
not far from those of drilling mud for which At = 0.33 is representative) of
$\rho^* = 1.3$ mass ratio of corn starch to (colored) water extruded from a
tube with inner diameter 12 mm.  A plunger was used to achieve a mass flow
rate and velocity of the suspension close to the gravity-driven flow rate of
water in the earlier experiment.  The Kelvin-Helmholtz instability was
suppressed (Fig.~1(b)), and the denser liquid descended as a coherent slug.

An instability analogous to the viscous buckling instability
\cite{T68,MK96,MRS98,D05} began at the leading edge of the flow, and led to
buckling and clumping into a slug of low aspect ratio.  The initial stage of
this instability is visible in Fig.~1(b) between the 100 cm and 110 cm
marks.  The viscous buckling instability occurs at low Re.  In our
experiments Re is large in the oil (but not well defined in the viscoelastic
corn starch suspension).  Instability is initiated by inertial forces at the
leading edge of the descending slug rather than by viscous forces, and its
growth is limited by the viscoelastic properties of the slug.  Fragmentation
of the descending column was only observed at flow speeds of 2.5 m/s, and
produced slugs of width comparable to the initial column radius.

The descent speed in high Re flow of a spherical slug whose size is that of
an oil well bore would be $\sim 2$ m/s.  A slug elongated vertically, as in
Fig.~1(b), would descend faster.  The Reynolds number in the oil is large
enough that the drag force is described by a coefficient of turbulent skin
friction $C_{skin} \approx 0.01$ \cite{SG00}, leading to a terminal descent
velocity $v_{descent} \approx \sqrt{\Delta \rho g r/(C_{skin} \rho)} \approx
7$ m/s for representative $\Delta \rho /\rho = 1$ and $r = 50$ mm.  This is
fast enough to overcome the upwelling even in an unusually rapidly flowing
well like Macondo.  In our experiments thick and vertically elongated
columns of dense viscoelastic fluid were affected by the inertial forces at
their leading surfaces; much longer columns and greater fluid masses would
be required to study this regime past the influence of the leading surface.

At much lower flow rates the denser aqueous viscoelastic fluid stretched
into a thin straight vertical filament, cohering viscoelastically, and
thinning as it was accelerated by gravity.  It did not disperse into drops
of smaller diameter.  In some trials, as illustrated in Fig.~1(c), it
remained straight and unbroken for $\gtrsim 1$ s, demonstrating suppression
of the Plateau-Rayleigh instability whose characteristic growth time
$\approx 3\sqrt{\rho r_{col}^3 / \sigma} \approx$ 7--20 ms for the column
radius $r_{col} \approx$ 0.5--1 mm \cite{dG02}.

\begin{figure}
\begin{center}
\includegraphics{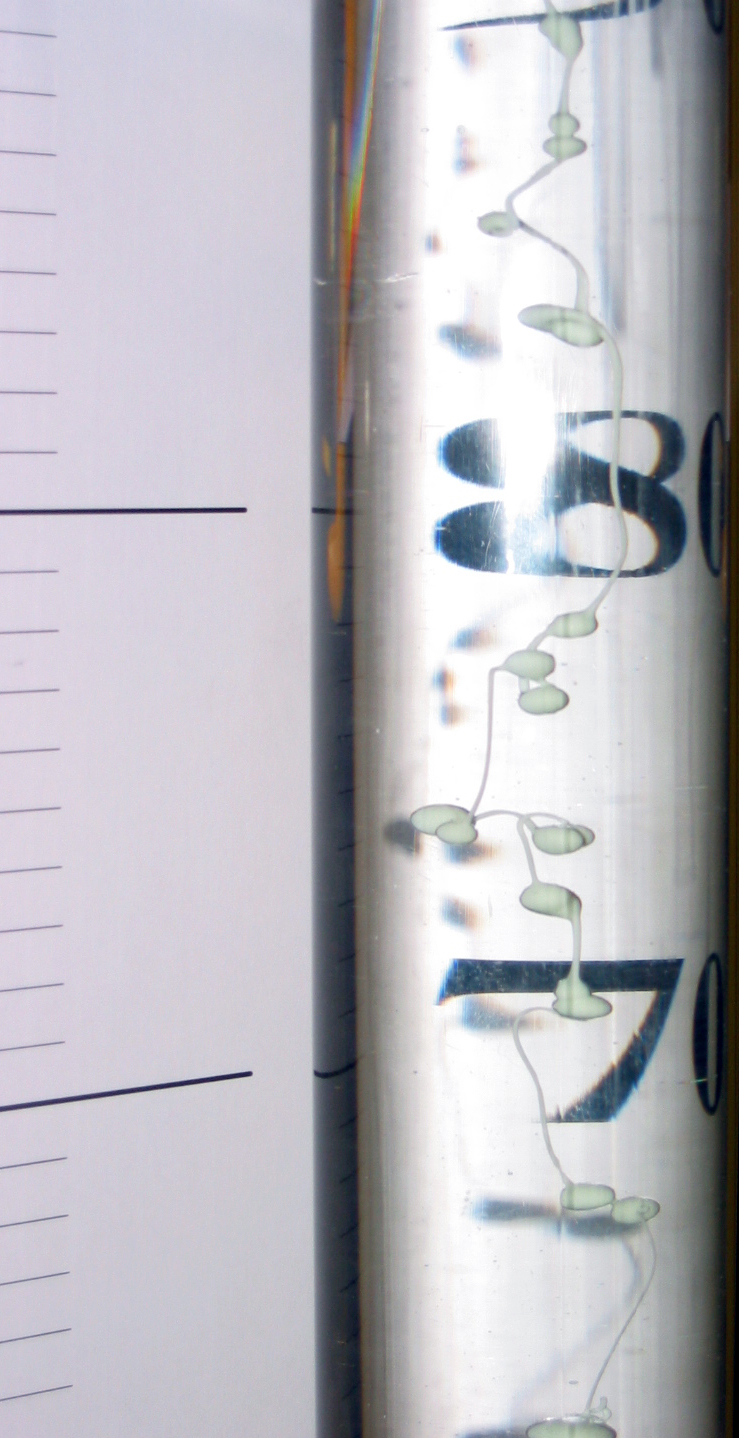}
\caption{Descent of $\rho^* = 1.2$ suspension at 1--2 ml/s showing globules
bound by loops of viscoelastic filament.  Vertical divisions are cm and tens
of cm.}
\label{globandloop}
\end{center}
\end{figure}

In order to study the transition between droplet formation (at $\rho^* = 0$) 
and continuous vertical flow ($\rho^* \geq 1.2$), we varied the suspension 
mass ratios and flow rates.  A number of complex phenomena were observed.
For example, at $\rho^* = 1.2$ and flow rates of 1--2 ml/s the leading
portion of the filament, after a steady descent of more than 1 m, broke up
into a complex flow of globules ($\approx 0.5$ cm in diameter) strung
together by thin filamentary loops, as shown in Fig.~\ref{globandloop}.
This appears to be the result of the combined effects of the unstable
dependence of flow rate on filament diameter (because of viscous drag by the
surrounding fluid, in analogy to the instability of resistive hydraulic flow
in open channels \cite{DP53}) at low Reynolds numbers, the viscous buckling
instability \cite{T68}, the Plateau-Rayleigh instability \cite{dG02}, and
viscoelasticity that inhibits breaking of the filaments even when they are
greatly stretched and thinned.

We conclude that the addition of dilatant polymers to make a viscoelastic
suspension is effective in suppressing both counterflow and surface
tension instabilities and may, making a Galilean transformation of reference
frames, maintain the coherence of a body of descending denser fluid in an
upwelling flow.  The potential practical application of these phenomena is
the use of viscoelastic dense mud to fill a flowing (blown-out) oil well by
enabling dense slugs of mud to descent against the upwelling oil.  The
hydrostatic pressure at the bottom may then be increased until it prevents
entry of additional oil from the reservoir.  This terminates the upwelling
and ``kills'' the well.  The phenomena observed go beyond this obvious
practical application and focus attention on the rich physics of two
counter-streaming fluids, one of which is shear-thickening and viscoelastic,
which has so far received little attention.  

\begin{acknowledgments}
We thank P.~Dimotakis and R.~Garwin for discussions that were the origin of
this experiment and R.~Grober for collaboration in a preliminary experiment
with miscible fluids.  This work was performed under the auspices of the
U.~S. DOE by LLNL under Contract DE-AC52-07NA27344 and Laboratory Directed
Research and Development project 10-FS-005.
\end{acknowledgments}

\end{document}